\documentclass[epj,nopacs]{svjour}
\usepackage{epsfig}
\usepackage{amssymb}
\usepackage{latexsym}
\newcommand{\HERMES}{{\sc Hermes}}

\begin{document}
\hugehead
\title{Measurement of Angular Distributions
and $R= \sigma_L/\sigma_T$
in
Diffractive Electroproduction of $\rho^0$ Mesons}

\author{ 
The HERMES Collaboration \medskip \\ 
K.~Ackerstaff$^{5}$,
A.~Airapetian$^{31}$,
N.~Akopov$^{31}$,
M.~Amarian$^{23,26,31}$,
E.C.~Aschenauer$^{6,12,23}$,
H.~Avakian$^{9,a}$,
R.~Avakian$^{31}$,
A.~Avetissian$^{31}$,
B.~Bains$^{14}$,
M.~Beckmann$^{11}$
% ...
} 

\institute{ 
$^1$Department of Physics, University of Alberta, Edmonton, Alberta T6G 2J1, 
Canada$^e$ \\
$^2$Physics Division, Argonne National Laboratory, Argonne, Illinois 60439-4843, 
USA$^f$ \\
$^3$W.K. Kellogg Radiation Laboratory, California Institute of Technology, Pasadena, California 91125, 
USA$^g$ \\
$^4$Nuclear Physics Laboratory, University of Colorado, Boulder, Colorado 80309-0446, 
USA$^h$\\
$^5$DESY, Deutsches Elektronen Synchrotron, 22603 Hamburg, Germany\\
$^6$DESY Zeuthen, 15738 Zeuthen, Germany\\
$^7$Joint Institute for Nuclear Research, 141980 Dubna, Russia\\
$^8$Physikalisches Institut, Universit\"at Erlangen-N\"urnberg, 91058 Erlangen, 
Germany$^{i,j}$\\
$^9$Istituto Nazionale di Fisica Nucleare, Laboratori Nazionali di Frascati, 00044 Frascati, Italy\\
$^{10}$Department of Physics, Florida International University, Miami, Florida 33199, USA \\
$^{11}$Fakult\"at f\"ur Physik, Universit\"at Freiburg, 79104 Freiburg, 
Germany$^i$ \\
$^{12}$Department of Subatomic and Radiation Physics, University of Gent, 9000 Gent, 
Belgium$^k$ \\
$^{13}$Max-Planck-Institut f\"ur Kernphysik, 69029 Heidelberg, Germany\\
$^{14}$Department of Physics, University of Illinois, Urbana, Illinois 61801, 
USA$^l$ \\
$^{15}$Department of Physics and Astronomy, University of Kentucky, Lexington, Kentucky 40506,USA \\
$^{16}$Physics Department, University of Liverpool, Liverpool L69 7ZE, 
United Kingdom$^m$ \\
$^{17}$Department of Physics, University of Wisconsin-Madison, Madison, Wisconsin 53706, 
USA$^n$ \\
$^{18}$Laboratory for Nuclear Science, Massachusetts Institute of Technology, Cambridge, Massachusetts 02139, 
USA$^o$ \\
$^{19}$Randall Laboratory of Physics, University of Michigan, Ann Arbor, Michigan 48109-1120, 
USA$^p$ \\
$^{20}$Lebedev Physical Institute, 117924 Moscow, Russia\\
$^{21}$Sektion Physik, Universit\"at M\"unchen, 85748 Garching, Germany$^i$ \\
$^{22}$Department of Physics, New Mexico State University, Las Cruces, New Mexico 88003, 
USA$^q$ \\
$^{23}$Nationaal Instituut voor Kernfysica en Hoge-Energiefysica (NIKHEF), 1009 DB Amsterdam, The 
Netherlands$^r$ \\
$^{24}$Department of Physics, University of Oregon, Eugene, Oregon 97403, USA\\
$^{25}$Petersburg Nuclear Physics Institute, St. Petersburg, Gatchina, 188350 Russia\\
$^{26}$Istituto Nazionale di Fisica Nucleare, Sezione Sanit\`a and Physics Laboratory, Istituto Superiore di Sanit\`a, 00161 Roma, Italy\\
$^{27}$Department of Physics, Simon Fraser University, Burnaby, British Columbia V5A 1S6, 
Canada$^e$\\
$^{28}$TRIUMF, Vancouver, British Columbia V6T 2A3, Canada$^e$ \\
$^{29}$Department of Physics, Tokyo Institute of Technology, Tokyo 152, 
Japan$^s$\\
$^{30}$Department of Physics and Astronomy, Vrije Universiteit, 1081 HV Amsterdam, The 
Netherlands$^r$ \\
$^{31}$Yerevan Physics Institute, 375036, Yerevan, Armenia\\
\footnoterule
$ ^a$ supported by INTAS contract No. 93-1827 \\
$ ^b$ supported by INTAS contract No. 1827-ext \\
$ ^c$ Deceased \\
$ ^d$ partially supported by the Thomas Jefferson National
Accelerator Facility, under DOE contract DE-AC05-84ER40150. \\
$ ^e$ supported by the Natural Sciences and Engineering Research 
Council of Canada (NSERC) \\
$ ^f$ supported by the US Department of Energy, Nuclear Physics Div.,
grant No. W-31-109-ENG-38 \\
$ ^g$ supported by the US National Science Foundation, grant No. PHY-9420470 \\
$ ^h$ supported by the US Department of Energy, Nuclear Physics Div.,
grant No. DE-FG03-95ER40913 \\
$ ^i$ supported by the Deutsche Bundesministerium f\"ur Bildung, 
Wissenschaft, Forschung und Technologie \\
$ ^j$ supported by the Deutsche Forschungsgemeinschaft \\
$ ^k$ supported by the FWO-Flanders, Belgium \\
$ ^l$ supported by the US National Science Foundation, grant No. PHY-9420787 \\
$ ^m$ supported by the U.K. Particle Physics and Astronomy Research Council \\
$ ^n$ supported by the US Department of Energy,  Nuclear Physics Div.,
grant No. DE-FG02-88ER40438, and the US National Science Foundation,
grant No. PHY-9722556 \\
$ ^o$ supported by the US Department of Energy, Nuclear Physics Div. \\
$ ^p$ supported by the US National Science Foundation, 
grant No. PHY-9724838 \\
$ ^q$ supported by the US Department of Energy, Nuclear Physics Div.,
grant No. DE-FG03-94ER40847 \\
$ ^r$ supported by the Dutch Foundation for Fundamenteel Onderzoek 
der Materie (FOM) \\
$ ^s$ supported by Monbusho, JSPS and Toray Science Foundation of Japan \\
} 

\date{Received: / Revised version:}

\titlerunning{Diffractive Electroproduction of $\rho^0$ Mesons}
\authorrunning{The HERMES Collaboration}

\abstract{Production and decay angular distributions were extracted from
measurements of exclusive electroproduction 
of the $\rho^0(770)$ meson over a range in the virtual photon 
negative four-momentum squared $0.5<Q^2<4$ GeV$^2$
and the photon-nucleon invariant mass range $3.8<W<6.5$ GeV.
The experiment was performed with the \HERMES\ spectrometer, using a
longitudinally polarized positron beam and a $^3$He gas target
internal to the HERA e$^\pm$ storage ring. 
The event sample combines $\rho^0$ mesons produced incoherently off individual
nucleons and coherently off the nucleus as a whole.  The distributions
in one production angle and two angles describing the $\rho^0
\rightarrow \pi^+\pi^-$ decay yielded measurements of eight elements
of the spin-density matrix, including one that had not been measured
before.  The results are consistent with the dominance of helicity
conserving amplitudes and natural parity exchange.  The improved
precision achieved at $4<W<7$ GeV, in combination with other data at
$W>7$ GeV, reveals evidence for an energy dependence in the
ratio $R$ of the longitudinal to transverse cross sections
at constant $Q^2$.} 
\maketitle

\section{Introduction}\label{sec:intro}

\begin{figure}[b]
\begin{center}
\epsfig{file=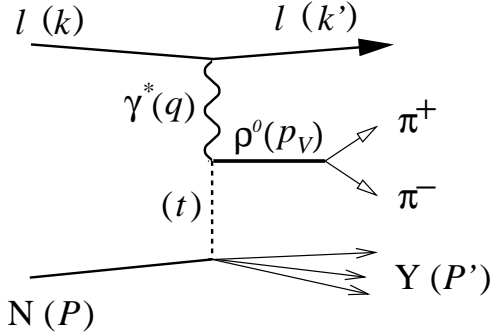,width=6.5cm}
\end{center}
\caption[dummy]{Tree level diagram of diffractive vector meson
electroproduction viewed as two-body virtual photoproduction.
Four-momentum labels are indicated in parenthesis.
}
\label{fig:VMD}
\end{figure}

  The study of exclusive photoproduction and leptoproduction
of vector mesons (leaving the target nucleon(s) intact)
has provided information both on the hadronic
components of the photon, and on the nature of diffractive scattering.
Theoretical models typically describe the 
exclusive production of light vector mesons
as occuring via the fluctuation of the real or virtual photon
into a quark-antiquark pair (or off-shell vector meson),
which is scattered onto the mass shell
by a diffractive interaction with the target.
The corresponding tree-level diagram is shown in Fig.~\ref{fig:VMD},
and a list of kinematic variables
describing the reaction is given in Table~\ref{tab:kin}.
The positron emits the virtual photon with energy
$\nu$ and four-mo\-men\-tum squared $-Q^2$, which then scatters from
the target by the exchange of particles in the $t$-channel.

\begin{table*}
\begin{center}
\caption{Kinematic variables used in the definition of exclusive
diffractive vector meson events. (Definitions involving energies or
three-vectors are given for the laboratory frame.)}
\label{tab:kin}
\begin{tabular}{ll}
\hline\noalign{\smallskip}
Variable                                &       Description     \\
\noalign{\smallskip}\hline\noalign{\smallskip}
$k=(E,{\vec{k}})$                  & 4-momentum of incident positron \\
$k'=(E',{\vec{k}}')$               & 4-momentum of scattered positron \\
$P=(M_{\mathrm{N}},\vec{0})$       & 4-momentum of target nucleon~N 
(ignoring Fermi motion and binding) \\
$q=(\nu,{\vec{q}})=k-k'$           & 4-momentum of virtual photon \\
$q^2=-Q^2$                         & Photon virtuality\\
$\theta_{\mathrm{e}}, \phi_{\mathrm{e}}$ & Positron scattering and azimuthal 
angles in the lab \\ & ($\phi_{\mathrm{e}}>0$ defines upper half of spectrometer)\\
$\epsilon=(1-\nu/E)/[1-\nu/E+\frac{1}{2}(\nu/E)^2]$
                                   & Virtual photon polarization parameter
 (valid for $m_{\mathrm{e}}^2 \ll Q^2 \ll E^2$)  \\
$W=|(q+P)|=(M_{\mathrm{N}}^2+2M_{\mathrm{N}}\nu-Q^2)^{\frac{1}{2}}$ &
                                       Total photon-nucleon 
centre-of-mass energy \\
$x=Q^2/2M_{\mathrm{N}}\nu$                    & Bjorken scaling variable \\
$y=\nu/E$                          & Fractional leptonic energy transfer \\
${\vec{p}}_{\mathrm{h}^{\pm}}$                & 3-momentum of positive or 
negative hadron \\ 
$E_{\pi^{\pm}}=\sqrt{m_{\pi}^2+ {\vec{p}}_{\mathrm{h}^{\pm}}^{\,2}}$
                                        & Hadron energy assuming it is a pion \\
$p_{\mathrm{V}}=(E_{\mathrm{V}},{\vec{p}_{\mathrm{V}}})=
(E_{\pi^+}+E_{\pi^-},
{\vec{p}_{\mathrm{h}^+}}
+
{\vec{p}_{\mathrm{h}^-}})$                  & Candidate $\rho^0$ 4-momentum \\
$M_{\pi\pi}=\sqrt{p_{\mathrm{V}}^2}$            & Candidate $\rho^0$ mass \\
$M_{\mathrm{Y}}=\sqrt{(P+q-p_{\mathrm{V}})^2}$                & Invariant mass of the recoiling baryonic 
system\\
$t=(q-p_{\mathrm{V}})^2<0$                         & Squared 4-momentum transfer to 
target\\
$t_0$                                   & Minimum of $|t|$ for fixed
$\nu$, $Q^2$, $M_{\pi\pi}$, and $M_{\mathrm{Y}}$ \\
$t'=t-t_0 < 0$                          & 4-momentum transfer squared relative 
to the case \\ 
                                    & where the meson and photon are collinear \\
$\Delta E=(M_{\mathrm{Y}}^2-M_{\mathrm{N}}^2)/2M_{\mathrm{N}}=
\nu-E_{\mathrm{V}}+t/2M_{\mathrm{N}}$ & Measure of energy absorbed by
recoil system \\

\noalign{\smallskip}\hline
\end{tabular}
\end{center}
\end{table*}

The decay-angle distribution is influenced by the transfer of the virtual 
photon's spin to that of the vector meson and can therefore provide a
stringent test of the reaction mechanism.
Electroproduction data are now available with modern precision
in the range of the photon-nucleon centre-of-mass energy $W$
reaching from low
($W<4\rm\,GeV$)~\cite{joos,bauer,delpapa,cassel},
through moderate ($6<W<20\rm\,GeV$)~\cite{chio,emc,nmc,e665},
and recently to very high ($50<W<140\rm\,GeV$)~\cite{zeus,zeusvio,h1} 
values,
and for values of
$Q^2$ up to 30\,GeV$^2$.
A variety of competing models mentioned below have achieved
some success in reproducing the observed dependences 
of the virtual photoproduction cross section
on $Q^2$, $W$, and $t$.
The polarization data presented here cover a range in $W$ that fills
the gap between 4 and 7 GeV
containing little previous data~\cite{dakin},
and should therefore serve as a useful additional constraint
on some of these models.

The Vector Meson Dominance (VMD) model~\cite{bauer} has 
traditionally been used to describe the reaction. It
treats a photon with $Q^2$ of order a few
GeV$^2$ or less as a superposition of the lightest
vector mesons that diffractively scatter from the target.
As $Q^2$ increases, 
the hadronic content
of the photon as a quark-antiquark pair is resolved.
More fundamental calculations including an explicit QCD-based treatment of the
dynamics of the $q\overline{q}$ pair provide a unified treatment
of the entire $Q^2$ range for either the total~\cite{pp0,nonper1}
or the longitudinal~\cite{pqcd3} cross section.

In both VMD~\cite{bauer} and
QCD-based models~\cite{nonper1,pom1,pom2},
the phenomenological Regge theory is used to describe the
diffractive reaction.
At low energy ($W<4\,$GeV) the
process is described in terms
of the exchange of mesons ($\rho$, $\omega$, $f_{2}$, $a_2$,
$\omega_3$, $\rho_3$,...), all lying on the same Regge trajectory~\cite{reggeon}.
This so-called Reggeon exchange describes the decrease with $W$
of the photoproduction cross section for real and quasi-real photons.
Above $W\approx 10\,$GeV, however,
the cross section increases slowly, and
Regge models of the reaction involve the
$t$-channel exchange of a colorless spinless object~\cite{nonper1,pom1,pom2},
which has been identified with the soft Pomeron introduced to
explain elastic pp scattering~\cite{pp0}.

The diffractive interaction has also been
treated in
perturbative QCD models of longitudinal $\rho^0$ production,
which illuminate how the Reggeons arise at
a fundamental level.
At the high $x$ values corresponding to low $W$ for moderate $Q^2$,
the diffractive interactions arise as the result
of the exchange of two quarks~\cite{guidal}, corresponding to
meson exchange in Regge theory.
At low $x$ (or high $W$), the interaction is assumed to involve
the exchange of two gluons~\cite{pqcd3},
which has led to attempts to
understand the phenomenological Pomeron in terms of nonperturbative exchange
of gluons~\cite{nonper2}.
At the highest $Q^2$, 
the divergence of the nucleon's gluon content at low $x$
leads to a strong $W$ dependence~\cite{pqcd3,pom2,pqcd1,pqcd2}.
At the intermediate energies ($3.8<W<6.5\rm\,GeV$) and photon virtualities 
($0.5<Q^2<4\rm\,GeV^2$) available to \HERMES, both
the quark exchange and the gluon exchange (or, in the Regge language,
both Reggeon and Pomeron exchange) processes are expected to
contribute.

For unpolarized targets and polarized beams, the distribution in
the production and decay angles
provides information on up to 26
independent elements of the $\rho^0$ spin density matrix~\cite{wolf},
providing much tighter constraints on theoretical models than
the total cross section alone.
Furthermore, the angular distribution yields important model-independent
insight into the reaction.
For example, experiments have indicated that
the helicity of the photon
in the $\gamma^* N$ centre-of-mass system
is approximately retained by the vector meson, a phenomenon
known as
$s$-channel helicity
conservation (SCHC)~\cite{bauer}.
Given the approximate validity of SCHC, the 
angular distribution
can be used to calculate the ratio
$R\equiv \sigma_L/\sigma_T$
of longitudinal to transverse
virtual-photon cross sections for exclusive $\rho^0$ meson production,
without performing
an experimentally difficult longitudinal-transverse (Rosenbluth) separation
using two different beam energies.
This ratio is expected to be especially sensitive to details of the
theoretical models.
For example, 
in ref.~\cite{lee2} it is asserted that $R(Q^2)$ is very sensitive to
nonperturbative effects in the longitudinal quark-photon coupling.
Recent calculations based on off-forward parton distributions have 
been used to describe the longitudinal cross section for
exclusive $\rho^0$-production~\cite{guidal,frankfurt,mankiewicz}.
Some features of these models
imply that in a complete QCD description higher-twist effects have to be
large, a  fact which should be kept in mind when interpreting the present data,
especially in view of the relatively low  $Q^2$-domain of \HERMES.

Phenomenological models are often simplified by the assumption of SCHC.
This assumption can be examined
in calculations of spin observables including a 
detailed treatment of the underlying quark-gluon degrees of freedom.
Violations of SCHC
could arise from gluon ladders and quark loops~\cite{svg}; moreover,
recent models of the quark-Po\-mer\-on coupling 
that address issues associated with gauge-invariance 
lead to non-zero spin-flip amplitudes~\cite{sk}.

This paper reports the results from the 1995 \HERMES\ data for the
angular distribution of $\rho^0(770)$ production and decay, 
from which eight matrix elements related to the $\rho^0$ spin density matrix 
were extracted. Some small matrix elements
that are sensitive to violations of SCHC are thereby better constrained 
in this energy region, 
including some that are accessible only using the longitudinal
polarisation of the lepton beam.
These data also provide the most precise determination of $R$ in this 
intermediate kinematic region containing only one previous 
data set~\cite{dakin}.
In combination with existing data at higher energy, these new data allow the 
$W$-dependence of $R$ to be investigated in the region 
above $W=4\,$GeV, excluding the lower region where 
there are indications that the character of the reaction
changes quickly. Below $W=4\,$GeV, 
the cross section increases rapidly with decreasing $W$,
and there appears evidence of significant resonance contributions
as well as a large phase difference between longitudinal
and transverse production that suggests the presence of non-diffractive
processes~\cite{joos}.

In Section \ref{sec:data}, the experimental conditions, event selection
and treatment of backgrounds are described.
The representation of the general decay-angular distribution in
terms of a set of matrix elements is presented in Section \ref{sec:struct}. 
The extraction of the matrix elements from the data 
is described in Section \ref{sec:extract}.
The results are presented and discussed in Section \ref{sec:results},
and the paper is summarized in Section \ref{sec:sum}.


\section{Experiment}\label{sec:data}

The experiment was performed using 
a gaseous $^3$He target contained in an open-ended cell coaxial
with the positron beam at the
HERA ep collider. During 1995 HERA produced a 27.5$\,$GeV positron
beam with an average longitudinal polarization of $0.48\pm 0.03$.
Luminosities up to $2{\times}10^{32}$ nucleons$\cdot$cm$^{-2}$s$^{-1}$
were achieved.
Scattered positrons and coincident hadrons were detected
in the \HERMES\ spectrometer, 
of which a detailed description can be found elsewhere~\cite{spectrometer}.
The spectrometer
is equipped with front and back drift chambers before and after 
a $1.3\,$T$\cdot$m dipole magnet. This tracking system provided a momentum
resolution $\Delta p/p\approx$\,1--2\% in the 1995 running period.
An iron plate that shields
the circulating beams from the bending-magnet field
separates the spectrometer into identically instrumented upper
and lower 
halves. 
The positron trigger
is produced by a timing hodoscope in coincidence with an
electromagnetic calorimeter
system; this system comprises a totally absorbing lead-glass
array and a preshower detector consisting of a passive
lead radiator and a scintillator hodoscope.
This trigger 
required a deposit of at least 4$\,$GeV
in the lead-glass array. 
A gas threshold {\v C}erenkov detector and a transition
radiation detector provided
additional positron/hadron separation.

\subsection{Selection of Events}

On-line hardware monitoring systems ensured that the only events considered
were those recorded when the
apparatus was fully operational.
Events with exactly one identified positron and two oppositely
charged hadrons were selected. 
Fiducial cuts were placed on the tracks to ensure that they all originated
from the target and fell
within the spectrometer acceptance.
Positrons and hadrons were distinguished with a likelihood analysis that
combined the response of the four particle-identification detectors
(see~\cite{spectrometer} and references therein).

\begin{figure}
\begin{center}
\epsfig{file=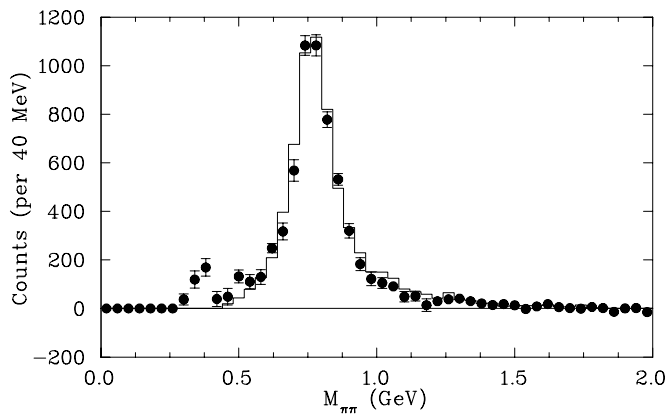,width=8.7cm}
\end{center}
\caption[dummy]{$M_{\pi\pi}$ distribution for exclusive diffractive events (filled
circles). The solid line is the yield from the Monte Carlo simulation,
discussed in
Section~\protect\ref{sec:mc}. The peak at low $M_{\pi\pi}$ is due to the 
$\mathrm{K^+K^-}$ decay mode of exclusively produced $\phi$ mesons 
(see section~\protect\ref{ss:bg}).}
\label{fig:plotm}
\end{figure}

\begin{figure}
\begin{center}
\epsfig{file=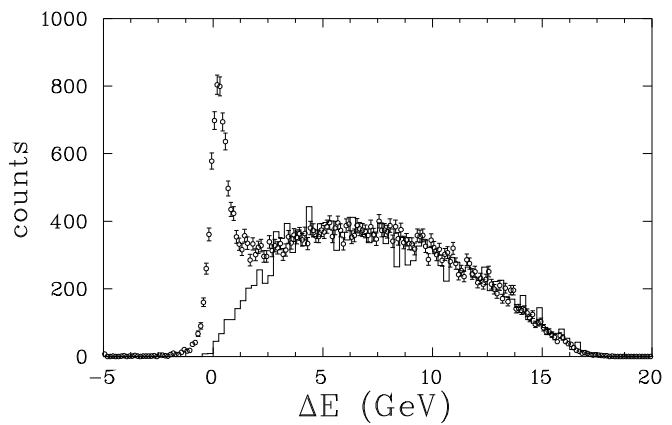,width=8.7cm}
\end{center}
\caption[dummy]{$\Delta E$ distribution for candidate diffractive
$\rho^0$ production events
(circles), showing the exclusive peak near zero. The histogram shows the
estimated contribution from non-diffractive processes
(see section~\ref{ss:nd}).}
\label{fig:delta_e}
\end{figure}

\begin{figure}
\begin{center}
\epsfig{file=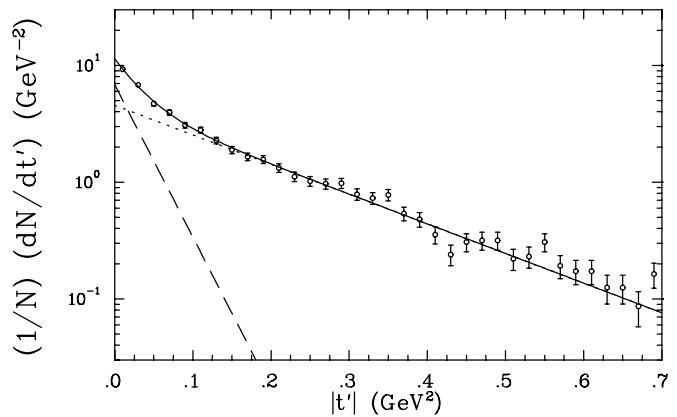,width=8.7cm}
\end{center}
\caption[dummy]{$t'$-distribution for exclusive $\rho^0$ production
events, with a fit (solid line) of (\ref{eq:dif}) to the data.
The contributions from coherent and incoherent scattering
are shown with dashed and dotted lines, respectively.}
\label{fig:tprime}
\end{figure}

The kinematics of the detected positron
define the four-momentum $q$ and polarization parameter $\epsilon$
of the virtual photon. The kinematic constraints applied during analysis are
\begin{eqnarray}
\nonumber
W   & >  & 3.8\,{\mathrm{GeV}}, \\
\nonumber
Q^2 & >  & 0.5\,{\mathrm{GeV}}^2, \\
\nonumber
 y   & <  & 0.85, \\ \nonumber
|\sin\theta_{\mathrm{e}}\sin\phi_{\mathrm{e}}| & > & 0.04\,.
\end{eqnarray}
The invariant mass $M_{\pi\pi}$ of the hadron pair was reconstructed 
under the assumption that both hadrons were pions (see Fig.~\ref{fig:plotm}).
The $\rho^0(770)$
mesons were identified by the requirement
$$
0.63<M_{\pi\pi}<0.91\,{\mathrm{GeV}}.
$$
The diffractive cross section is forward peaked for small 
$t'\simeq -p_\perp (\rho^0)$ 
(see Table~\ref{tab:kin}):
\begin{equation}
\nonumber
\frac{d\sigma}{dt'}\propto\,e^{bt'}.
\end{equation}
Here the slope parameter $b$ is related to the RMS radii 
$r_{\rho}$ and $r_{\mathrm{N}}$ of
the meson and nucleon or nucleus~\cite{povh}:
\begin{equation}
\nonumber
b\approx \frac{1}{3}(r_{\rho}^2 +r_{\mathrm{N}}^2).
\end{equation}
Hence diffractive events are generally characterized by small values of $t'$
leaving the gently recoiling target remnant well separated
from the forward-going vector meson.
The acceptance of
the \HERMES\ spectrometer precludes detection of the recoiling
target or target remnant; however, because of the small energy width of the
HERA positron
beam ($\sigma_E\approx 25\,$MeV) and the good momentum and angular
resolutions of
the \HERMES\ spectrometer, the energy $\Delta E$ absorbed by the
undetected target remnant can be reconstructed (see Table~\ref{tab:kin} and
Fig.~\ref{fig:delta_e}) and used
to reject events in which the target nucleon(s) do not remain intact.
Exclusive
diffractive events were consequently selected by adding the kinematic requirements
\begin{eqnarray}
\Delta E  & < 0.7\,{\mathrm{GeV}}, \label{eq:decut} \\
\nonumber
-t' &  <  0.5\,{\mathrm{GeV^2}}.  
\end{eqnarray}
Both coherent scattering from $^3$He and incoherent scattering from
nucleons are contained in the peak at $\Delta E\approx 0$, though
the peaking in $t'$ is stronger for coherent scattering because of the
larger size of the $^3$He nucleus compared with the nucleon
(Fig.~\ref{fig:tprime}). 
Both contributions were included in the present analysis.
The above experimental requirements were applied, to the extent applicable,
in all of the results reported here, resulting in 4800 accepted events.

\subsection{Corrections for Background}\label{ss:bg}
 Background arises from both diffractive and non-diff\-rac\-tive processes.
Inelastic fragmentation, {\it i.e.}~hard scattering on partons followed
by hadronization, is the primary source of
non-diffractive background.
The diffractive backgrounds include
exclusive non-resonant $\pi^+\pi^-$
production in which
the pions are produced without
forming a resonance, double-diffractive dissociation in which
the target remnant breaks up during diffractive vector-meson
production, and exclusive production of $\omega$ mesons.
The $\mathrm{K^+K^-}$ decay mode of the $\phi$ meson is reconstructed 
with $M_{\pi\pi}$ in the region below 0.63\,GeV
when pion masses are assumed for the kaons, and is thus excluded from
the selected data sample.
The pair production ($\gamma\rightarrow
\mathrm{e^+e^-}$) contamination of the di-hadron sample 
resulting from mis-identification of the leptons is negligible.

\subsubsection{Non-diffractive Background}\label{ss:nd}
Non-diffractive processes
constitute the dominant source of background,
for which a correction was applied using a Monte Carlo simulation.
Background events were generated according to
the non-diffractive deep-inelastic cross section using LEPTO,
and the fragmentation was treated using the LUND model, both described
in ref.~\cite{lepto6.5}. The detector
response was simulated with a GEANT-based Monte Carlo code~\cite{spectrometer}.
Three-track $\mathrm{eN\rightarrow eh^+h^-X}$ events were
selected from these Monte Carlo events and
subjected to
the same analysis criteria as the data, thereby producing a sample
of non-diffractive background events.
The Monte Carlo estimation was normalized to the data in
the inelastic region defined by $\Delta E>4\,$GeV
(see Fig.~\ref{fig:delta_e}). The experimental yields for $\Delta E<0.7\,$GeV
were corrected 
by 6\% at small $Q^2$ to 15\% at large $Q^2$,
using this sample of background events. 

\subsubsection{Diffractive Background}
\begin{figure}[t]
\begin{center}
\epsfig{file=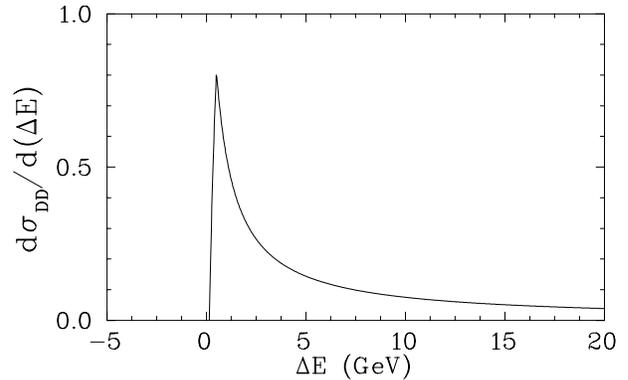,width=8cm}
\end{center}
\caption{A parameterization of the target excitation spectrum of 
double-diffractive dissociation (arbitrary units).}
\label{fig:dd}
\end{figure}

        Measurements of double-diffractive pp scattering have suggested
that the mass spectrum of the baryonic system ($Y$) resulting
from dissociation of the target has the form~\cite{e665,pp}
\[ M_{\mathrm{Y}}^2 { {d \sigma} \over {d M_{\mathrm{Y}}^2} }
\propto \left\{ \begin{array}{ll}      
         {{M_{\mathrm{Y}}^2 - (M_{\mathrm{N}} + m_{\pi})^2}\over 
         {1.8\,{\rm GeV}^2}-(M_{\mathrm{N}} + m_{\pi})^2 }
                   & \mbox{if $M_{\mathrm{Y}}^2 < $ 1.8$\,$GeV$^2$} \\
              1    & \mbox{otherwise.}
                 \end{array}
          \right. \]
This function is shown in Fig.~\ref{fig:dd} expressed in terms of 
$\Delta E=(M_{\mathrm{Y}}^2-M_{\mathrm{N}}^2)/2M_{\mathrm{N}}$.  
Only 15\% of diffractive events with target dissociation 
described by such a model
would pass the $\Delta E$ cut used in this analysis,
which is a benefit of the good energy resolution compared
to that available in previous fixed-target leptoproduction experiments
cited herein.
Recent measurements
of diffractive $\rho^0$ production with target dissociation 
at centre-of-mass energies from 60$\,$GeV to 180$\,$GeV give
$\sigma_{\gamma^*p \rightarrow \rho^0 Y}/
\sigma_{\gamma^*p \rightarrow \rho^0 p}=0.65\pm0.17$,
with a diffractive slope parameter $b_{DD}=2.1\pm 0.7$\,GeV$^{-2}$~\cite{ddh1}.
Assuming that the \HERMES\ acceptance is unaffected by target dissociation,
this result implies that the contamination
of the exclusive data by double-diffractive events is less than ($6\pm 2$)\%.
Moreover, no difference in the measured longitudinal fraction of $\rho^0$ mesons
was seen between the two processes~\cite{ddh1}. 
This is not surprising as target
dissociation is not expected to affect the helicity structure of the
$\gamma^*\rho^0$ vertex. Hence no correction or uncertainty
was applied for this source of background since this contribution to the
uncertainty is expected to be small compared to others.

        The measured distribution includes interfering contributions from
both diffractive production of 
$\omega(783)\rightarrow\pi^+\pi^-$ mesons (branching ratio 2.2\%) 
and non-resonant exclusive production of $\pi^+\pi^-$ pairs, neither of which
are distinguished here from
diffractive $\rho^0$ production~\cite{ryskin1}.
However, the $\omega$ production cross section is down by an order of magnitude,
and its contribution tends to be cancelled by 
the enhancement of the background subtraction induced by
the contribution to the region at larger $\Delta E$ 
by the dominant decay branch $\omega \rightarrow \pi^+\pi^-\pi^0$. 
In principle, the small non-resonant contribution should be included in any 
theoretical treatment that is compared to the data.

\subsection{Radiative Effects}
  Calculated radiative corrections to the lowest order diagram 
(Fig.~\ref{fig:VMD}) have approximately a 20\% effect on the measured
exclusive vector meson production cross section.
Radiative effects on polarization phenomena are typically much smaller.
External radiative effects in the spectrometer are entirely accounted for 
in the GEANT-based Monte Carlo simulation.
Because of the negligible target thickness and windowless cell,
external bremsstrahlung before scattering is negligible.
No corrections were applied for internal radiative effects; they are mitigated by
the cut on $\Delta E$ (\ref{eq:decut}), which ensures that no hard photon
is radiated in the events used to reconstruct the production and decay angles.
A systematic uncertainty was assigned for internal radiation, 
as discussed below. 

\section{Angular Distribution of the Production and Decay}\label{sec:struct}
\subsection{The $s$-Channel Helicity Frame}
\begin{figure}
\begin{center}
\epsfig{file=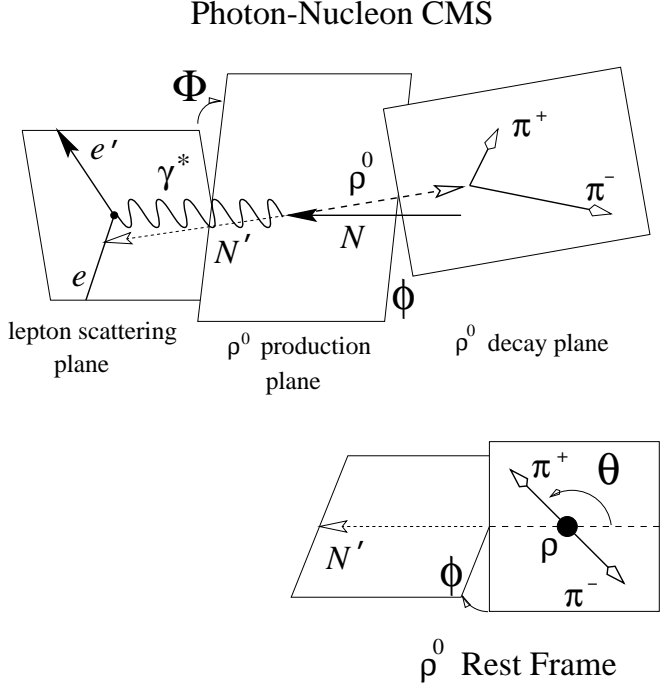,width=8.7cm}
\end{center}
\caption{The $s$-channel helicity frame.}
\label{fig:hellen}
\end{figure}

Analyses of previous data on exclusive vector-meson
production have shown
that only the $s$-channel helicity frame
leads to matrix elements
that are not strongly $t$-de\-pend\-ent~\cite{crittenden,ballman}.
The $s$-channel helicity frame is the
centre-of-mass system of the $\gamma^* N$ system, 
with the quantization axis 
of the $\rho^0$ 
opposite to
the direction of the recoiling target.
The reconstructed kinematics are essentially indistinguishable
for coherent and incoherent scattering, corresponding to
the use of the nuclear or nucleon mass.
The azimuthal production angle
$\Phi$ (see Fig.~\ref{fig:hellen}) is defined in
this frame as the angle
between the lepton scattering plane and the
$\rho^0$ production plane containing the momentum transfer to the target.
The decay-angle distribution can then be expressed in the $\rho^0$ rest frame
in terms of the
polar and azimuthal decay angles $\theta$ and $\phi$  relative to the
same quantization axis,
with the azimuth defined relative to the vector meson production plane
(see Fig.~\ref{fig:hellen}).
The angular distribution ${\cal W}(\cos\theta,\phi,\Phi)$
is then completely characterized by the three angles
$\theta$, $\phi$, and $\Phi$. The spin state of the $\rho^0$
is reflected in the orbital angular momentum of the pion pair,
giving rise to the $(\cos\theta,\phi)$ dependence.
The $\Phi$ dependence of ${\cal W}$ arises because the virtual photon
polarization depends on $\Phi$.

\subsection{Decomposition of the Angular Distribution}

 A complete derivation of the decay-angle distribution can be found
in ref.~\cite{wolf}. Briefly,
the decay distribution can be related to the polarization
density matrix $\rho_{\lambda\lambda'}$ of the vector meson
expressed in the helicity basis,
using the Wigner $D$-functions:
\begin{eqnarray}
\nonumber
\lefteqn{{\cal W}(\cos\theta,\phi,\Phi)=} \\ 
& \frac{3}{4\pi}
{\displaystyle \sum_{\lambda\lambda'=-1,0,1}}
D^{1\,*}_{\lambda,0}(\phi,\theta,-\phi)
\rho_{\lambda\lambda'}(\Phi)
D^1_{\lambda',0}(\phi,\theta,-\phi). 
& ~~~~ 
\label{eq:wigner}
\end{eqnarray}
The $\Phi$ dependence appears only in $\rho_{\lambda\lambda'}$,
which can be decomposed into 9 independent contributions 
characterized by the matrix elements
$\rho^\alpha_{\lambda\lambda'}$:
$\alpha=0,1,2,3$ represent transverse photons:
unpolarized, the two directions of linear polarization, and
circular polarization, respectively.
Pure longitudinal photons correspond to $\alpha=4$.
The remaining matrices $\alpha=5$\ldots$8$ are attributable to
interference of longitudinal with transverse photons.
Expressions can be found in Appendix~A of ref.~\cite{wolf}
giving the various $\rho^{\alpha}_{\lambda\lambda'}$
in terms of combinations of the Wick amplitudes $T_{\lambda\gamma}$
for a photon of helicity $\gamma$ to produce a meson of
helicity $\lambda$, summed over nucleon spins.

At fixed beam energy, it is not possible to
explicitly separate the contributions from longitudinal and transverse
photons, and the angular distributions are described in terms of the
following linear combinations of the $\rho^{\alpha}_{\lambda\lambda'}$:
\begin{eqnarray}
\nonumber
r^{04}_{\lambda\lambda'}& = &{{\rho^0_{\lambda\lambda'}+\epsilon R \rho^4_{\lambda\lambda'}}\over{1+\epsilon R}},\\
\nonumber
r^{\alpha}_{\lambda\lambda'}& =&
{{\rho^{\alpha}_{\lambda\lambda'}}\over{1+\epsilon R}},\ \ \ \ \alpha=(1,2,3), \\
\nonumber
r^{\alpha}_{\lambda\lambda'}& =&
\sqrt{R}{{\rho^{\alpha}_{\lambda\lambda'}}\over{1+\epsilon R}},\ \ \alpha=(5,6,7,8).
\end{eqnarray}
Here $R$ is the ratio of longitudinal to transverse cross sections
and $\epsilon$ is the polarization parameter of the virtual photon. 
Equation~(\ref{eq:wigner}) becomes~\cite{wolf}
\begin{eqnarray}
\nonumber
{\cal W}(\cos\theta,\phi,\Phi)&=&
{{3}\over{4\pi}}
\Big\{
 {{1}\over{2}} (1-r^{04}_{00})
+{{1}\over{2}} (3r^{04}_{00}-1)\cos^2\theta \\
\nonumber
& &
\hspace{-1cm}
-\sqrt{2}\,\Re(r^{04}_{10})\sin 2\theta \cos\phi 
-r^{04}_{1-1}\sin^2\theta\cos 2\phi \\
\nonumber
& &
\hspace{-2.6cm}
-\epsilon \cos 2\Phi
 \Big(r^{1}_{11}\sin^2\theta +r^{1}_{00}\cos^2\theta
        -\sqrt{2}\,\Re(r^{1}_{10})\sin 2\theta\cos\phi \\
\nonumber
& &
        -r^{1}_{1-1}\sin^2\theta\cos 2\phi \Big) \\
\nonumber
& &
\hspace{-2.6cm}
-\epsilon \sin 2\Phi
 \left(\sqrt{2}\,\Im(r^{2}_{10})\sin 2\theta\sin\phi  
        + \Im(r^{2}_{1-1})\sin^2\theta \sin 2\phi\right) \\
\nonumber
& &
\hspace{-2.6cm}
+\sqrt{2\epsilon(1+\epsilon)}\cos\Phi
\Big(r^5_{11}\sin^2\theta+r^5_{00}\cos^2\theta \\
\nonumber
& &
\hspace{-1cm}
        -\sqrt{2}\,\Re(r^5_{10})\sin 2\theta\cos\phi
        -r^5_{1-1}\sin^2\theta\cos 2\phi \Big) \\
\nonumber
& &
\hspace{-2.6cm}
+\sqrt{2\epsilon(1+\epsilon)}\sin\Phi \times \\
\nonumber
& &
\hspace{-1cm}
\left(\sqrt{2}\,\Im(r^{6}_{10})\sin 2\theta\sin\phi
        +\Im(r^{6}_{1-1})\sin^2\theta\sin 2\phi\right)\\
\nonumber
& &
\hspace{-2.6cm}+P_b\Big[\sqrt{1-\epsilon^2} \times \\
\nonumber
& &
\hspace{-1cm}\left(\sqrt{2}\,\Im(r^{3}_{10})\sin 2\theta\sin \phi 
        +\Im(r^{3}_{1-1})\sin^2 \theta\sin 2\phi\right)\\
\nonumber
& &
\hspace{-2cm}
+\sqrt{2\epsilon(1-\epsilon)}\cos\Phi \times \\
\nonumber
& &
\hspace{-1cm}
\left(\sqrt{2}\,\Im(r^{7}_{10})\sin 2\theta\sin \phi
        +\Im(r^{7}_{1-1})\sin^2\theta\sin 2\phi\right) \\
\nonumber
& &
\hspace{-2cm}
+\sqrt{2\epsilon(1-\epsilon)}\sin\Phi \times \\
\nonumber
& &
\hspace{-1cm}
\Big(r^{8}_{11}\sin^2\theta +r^{8}_{00}\cos^2\theta
        -\sqrt{2}\,\Re(r^{8}_{10})\sin 2\theta\cos\phi \\
& &
        -r^{8}_{1-1}\sin^2\theta\cos 2\phi\Big)
\Big]\Big\},
\label{eq:w_full}
\end{eqnarray}
where $P_b$ is the longitudinal polarization of the lepton beam,
and the approximation $Q^2\gg m^2_{\mathrm{e}}$
has been made.

The statistical precision of the present data set
required that the
analysis be limited to only one-dimensional angular distributions.

\subsubsection{The $\bf{\cos\theta}$ Distribution}
 Integrating (\ref{eq:w_full}) over $\phi$ and averaging over $\Phi$
gives
\begin{equation}
{\cal W}(\cos\theta)= {{3}\over{4}}\left[
1- r^{04}_{00} +(3r^{04}_{00} -1){\cos}^2 \theta\right].
\label{eq:cos}
\end{equation}
This distribution depends on only one matrix element $r^{04}_{00}$, which
can be identified as the longitudinal fraction of $\rho^0$ mesons.
${\cal W}(\cos\theta)$ is not constrained by SCHC.

\subsubsection{ The $\psi$ Distribution}

The polarization angle $\psi\equiv\phi-\Phi$
is the angle between the decay plane of the $\rho^0$ and the plane of
polarization of
the photon (i.e. the lepton scattering plane).
If SCHC is valid, the azimuthal dependence
is a function of $\psi$ only, 
and integration of (\ref{eq:w_full})
over $\cos\theta$ gives
\begin{equation}
\label{eq:psi}
   {\cal W}(\psi) = {{1}\over{2\pi}}\left[1+2\epsilon r^{1}_{1-1} \cos 2\psi\right],
\end{equation}
which can be interpreted as an interference pattern
arising from the amplitudes for production 
by the two linear polarization states of the photon.
A positive value for $r^{1}_{1-1}$ indicates that the meson
and photon spins are aligned such that
decay pions are preferentially emitted in the lepton scattering plane.

\subsubsection{The $\bf{\phi}$ Distribution}
 Integrating (\ref{eq:w_full})
over $\cos\theta$ and averaging over $\Phi$ gives
\begin{eqnarray}
\nonumber
\lefteqn{{\cal W}(\phi) =}\\ 
& {{1}\over{2\pi}}\left[1-2r^{04}_{1-1}\cos 2\phi +
P_b\sqrt{1-\epsilon^2}\,\Im(r^{3}_{1-1})
\sin{2\phi}\right]. 
& ~~~
\label{eq:phi}
\end{eqnarray}
The azimuthal decay angle $\phi$ is sensitive to the linear polarization
of the $\rho^0$ and (\ref{eq:wigner}) produces only terms of the form
$e^{i(\lambda-\lambda')\phi}$ with $|\lambda-\lambda'|=2$.
This distribution measures any SCHC-violating
couplings of the linear 
polarization ($\lambda=1,\lambda'=-1$)
of the vector meson to transverse
unpolarized\,($\alpha=0$) and circularly polarized\,($\alpha=3$),
and longitudinal\,($\alpha=4$) photons.
If SCHC holds, then ${\cal W}(\phi)=1/2\pi$.

\subsubsection{ The $\Phi$ Distribution}
Integration of ${\cal W}(\cos\theta,\phi,\Phi)$ over 
$\cos\theta$ and averaging over $\phi$ 
is equivalent to summing over the final state
helicities ($\lambda,\lambda'$), giving
\begin{eqnarray}
\nonumber
{\cal W}(\Phi)&=&{{1}\over{2\pi}}
\left\{\right.
1-\epsilon\cos 2\Phi\,{\mathrm{Tr}}(r^1) 
+\sqrt{2\epsilon(1+\epsilon)}\cos\Phi\,{\mathrm{Tr}}(r^5) \\ 
 & &
+P_b\sqrt{2\epsilon(1-\epsilon)}
\sin\Phi\,{\mathrm{Tr}}(r^8)
\left.\right\}.
\label{eq:Phi}
\end{eqnarray}
As a sum over the $\rho^0$ decay states,
${\cal W}(\Phi)$ measures
the out-of-plane dependence of the $eN\rightarrow eN\rho^0 $ reaction,
and the matrix elements
$\mathrm{Tr}(r^1)$, $\mathrm{Tr}(r^5)$,
and $\mathrm{Tr}(r^8)$ can be identified with the out-of-plane response
functions familiar from medium-energy 
quasi-elastic electron scattering: $W_{TT}$, $W_{LT}$, and $W_{LT'}$,
respectively (see~\cite{vman}). 
Parity conservation ensures that
$r^{\alpha}_{11}=r^{\alpha}_{-1-1}$, for $\alpha=0,1,4,5,8$; thus
${\mathrm{Tr}}(r^{\alpha})=2r^{\alpha}_{11}+r^{\alpha}_{00}$, in
agreement with the notation of ref.~\cite{wolf}.
If SCHC applies, then ${\cal W}(\Phi)=1/2\pi$.

\section{Extraction of Matrix Elements}\label{sec:extract}

To account for radiative and instrumental effects on the measurements,
a Monte Carlo event generator for exclusive diffractive $\rho^0$ production 
was developed and used in conjunction with the GEANT detector simulation.
The matrix elements were extracted as those values that best fitted 
the Monte Carlo distributions to the data. 

\subsection{Monte Carlo Event Generator}
\label{sec:mc}

The diffractive $\rho^0$ generator produces
$eN\rightarrow eN\rho^0 $
events according to the leptoproduction double-differential 
cross section~\cite{bauer}
\begin{equation}
{{d\sigma}\over{dQ^2dE'}}=\frac{1}{4\pi EE'}\Gamma_T(Q^2,E')\sigma(Q^2),
\nonumber
\end{equation}
where the virtual photon flux is defined 
according to the Hand convention~\cite{hand}
\begin{equation}
\Gamma_T(Q^2,E')=
{{\alpha}\over{4\pi^2}}
{{W^2-M_{\mathrm{N}}^2}\over{2M_{\mathrm{N}}}}
{{1}\over{Q^2}}
{{E'}\over{E}}
{{2}\over{1-\epsilon}}.
\nonumber
\end{equation}
Using VMD, the virtual photon
production cross section can be expressed in terms of the
real photoproduction cross section as~\cite{bauer,cassel}
\begin{eqnarray}
\sigma(Q^2)     &=&\frac{1}{(1+Q^2/M^2_{\rho})^2}\cdot\sigma(Q^2=0), 
\label{eq:q2} \\
\sigma(Q^2=0)   &=&A_{\gamma}\,\frac{2M_{\mathrm{p}}}
                                    {W^2-M^2_{\mathrm{p}}} +B_{\gamma}, 
\label{eq:cassel} 
\end{eqnarray}
with
$A_{\gamma}=29.4\,\mu{\mathrm{b}}\cdot{\mathrm{GeV}}$ and
$B_{\gamma}=9.5\,\mu{\mathrm{b}}$ from a fit to data from ref.~\cite{cassel}.

The $\rho^0$ mass selection was weighted according to a skewed Breit-Wigner
distribution
\begin{equation}
\frac{dN}{dM_{\pi\pi}}=
{{M_{\pi\pi}\Gamma_{\rho} M_{\rho}}
\over{(M_{\pi\pi}^2-M_{\rho}^2)^2+M^2_{\rho}\Gamma_{\rho}^2}}
\left( {{M_{\rho}}\over{M_{\pi\pi}}}\right)^{n_{\mathrm{s}}}.
\label{eq:BW}
\end{equation}
The mass-dependent width $\Gamma_{\rho}(M_{\pi\pi})$ and skewing parameter
$n_{\mbox{s}}=3.18$ were taken from ref.~\cite{skew}.

The events were generated from an exponential distribution in $t'$,
and weighted to account
for the coherent contribution at low $t'$:
\begin{equation}
\frac{dN}{dt'}\propto
\frac{\sigma_{\mathrm{A}}}{\sigma_{\mathrm{N}}}b_{\mathrm{A}} 
                           e^{b_{\mathrm{A}}t'}+b_{\mathrm{N}} 
                                        e^{b_{\mathrm{N}}t'}.
\label{eq:dif}
\end{equation}
The ratio of the total coherent cross section from $^3$He to the 
incoherent one from nucleons was set to 
$\sigma_{\mathrm{A}}/\sigma_{\mathrm{N}}=0.48$,
and the diffractive slope for $^3$He (the nucleon) was taken as
25 (6)$\,$GeV$^{-2}$ ({\it e.g.}\ see~\cite{bauer}).
The present data set was found to be consistent with 
(\ref{eq:q2}--\ref{eq:dif}) using the parameters 
as given above. While the value of the skewing parameter
$n_{\mbox{s}}=3.0\pm0.02$ provides a better fit~\cite{Kolst}, this hardly
affects the angular distributions. 

\subsection{Procedure}

The analysis was done in four $Q^2$ regions
containing data of approximately equal statistical precision:
0.5 \ldots\ 0.95 \ldots\ 1.3 \ldots\ 1.96 \ldots\ 4~GeV$^2$.
For each $Q^2$ bin, a sample of Monte Carlo events 
was generated from a distribution that is isotropic in all angles.
These events were then subjected to the simulation of the experiment and
normal data analysis procedures, accounting for the rather severe effects
of the spectrometer acceptance.
In order to impose the effects of the $r^\alpha_{\lambda\lambda'}$ 
on the shapes of the simulated
angular distributions, the surviving $N_{MC}\approx 1.5\times 10^4$ events 
were weighted according to
\begin{eqnarray}
\nonumber
w_i={\cal W}(\cos\theta_i, \phi_i, \Phi_i; \epsilon_i) &   & (i=1,...,N_{MC}),
\end{eqnarray}
with ${\cal W}(\cos\theta_i, \phi_i, \Phi_i;\epsilon_i)$ given by 
(\ref{eq:w_full}).
The $\cos\theta_i$, $\phi_i$, $\Phi_i$, and $\epsilon_i$ refer to
the generated kinematics of each Monte Carlo event. The weights $w_i$
were evaluated as
a function of the 23 $r$-matrix elements describing ${\cal W}(\cos\theta,\phi,\Phi)$
(\ref{eq:w_full}) as follows:
\begin{itemize}
\item The matrix elements that control the one-dimensional distributions
${\cal W}(\cos\theta)$, ${\cal W}(\phi)$, ${\cal W}(\Phi)$, and ${\cal W}(\psi)$,
were taken as free parameters:
$r^{04}_{00}$,
$r^{04}_{1-1}$, 
$r^1_{1-1}$, 
$r^{3}_{1-1}$, 
${\mathrm{Tr}}\,(r^1)$,
${\mathrm{Tr}}\,(r^5)$ and
${\mathrm{Tr}}\,(r^8)$.
\item An eighth free parameter was taken to be the phase $\delta$ 
between the longitudinal and transverse Wick amplitudes $T_{00}$ and $T_{11}$ ---
the only ones remaining under the assumption of SCHC and
natural parity exchange~\cite{wolf}.
This parameter was used to constrain the following matrix elements:
\begin{eqnarray}
-\Im(r^6_{10})&=&\Re(r^5_{10})\equiv\frac{1}
{2}\sqrt{\frac{R}{2}} \frac{\cos{\delta}}{(1+\epsilon R)}, 
\label{eq:constrain2} \\
\Im(r^7_{10})&=&\Re(r^8_{10})\equiv\frac{1}
{2}\sqrt{\frac{R}{2}} \frac{\sin{\delta}}{(1+\epsilon R)},
\label{eq:constrain1}
\end{eqnarray}
with $R$ calculated from $r^{04}_{00}$ and $\epsilon$
according to (\ref{eq:rlt}). 
The $r^{5,6,7,8}_{10}$ terms, due to longitudinal-transverse
interference, cause a correlation between $\cos\theta$ and $\psi$
and are not observable in one-dimensional angular distributions; however,
the correlation between $\cos\theta$ and $\phi$ in the detector acceptance 
function caused by the intrusion of the magnet shielding plate
requires that all matrix elements must be fitted simultaneously.
\item The matrix elements describing the azimuthal production angle ($\Phi$)
distribution were assumed to have the helicity structure
\begin{eqnarray}
r^{1,5,8}_{00}&=&{\mathrm{Tr}}\,(r^{1,5,8})\ r^{04}_{00},
\label{eq:Phi00} \\
r^{1,5,8}_{11}&=&{\mathrm{Tr}}\,(r^{1,5,8})\ \frac{1}{2}(1-r^{04}_{00}),
\label{eq:Phi11}
\end{eqnarray}
which assigns to these 
amplitudes the same ratio of longitudinal-to-transverse strength
as observed in the distribution averaged over $\Phi$.
\item The remaining matrix elements are not observable in the one-dimensional
distributions and were constrained assuming SCHC~\cite{wolf} by
\begin{eqnarray}
\nonumber
\Im(r^2_{1-1})&=&-r^1_{1-1}, \\
\nonumber
r^{5,6,7,8}_{1-1} & = & 0, \\
\nonumber
r^{1,2,3,04}_{10} & = & 0.
\end{eqnarray}
\end{itemize}

The 8 free parameters plus a single arbitrary global normalization factor 
were simultaneously adjusted to minimize the
total $\chi^2$ of the deviations between the data and the
Monte Carlo model in their reconstructed 
yield distributions $dN/d\cos\theta$,
$dN/d\phi$,
$dN/d\Phi$,
and $dN/d\psi$. 
The $\chi^2$ per degree of freedom ranges from 29/42 to 48/42
for the four $Q^2$ bins.
The optimized parameters and their
uncertainties are taken to be the measured values at each $Q^2$ point. 
The quoted statistical uncertainties include the effects of
the significant correlations between parameters.

\subsection{Systematic Uncertainties}
Systematic uncertainties due to assumptions used in the analysis 
were estimated by
performing the analysis under different yet tenable assumptions
and comparing results. The following issues were investigated:
\begin{itemize}
\item Geometric acceptance:
The systematic uncertainty due to alignment errors was estimated
by varying the alignment in the GEANT model within known tolerances
and performing the analysis procedure on Monte Carlo data.
This contribution tends to dominate at small $Q^2$ for all matrix
elements except Tr($r^1$) and Tr($r^8$).
\item Assumptions in the VMD-based diffractive generator:
$Q^2$ and $W$ dependence ((\ref{eq:q2}) and (\ref{eq:cassel})),
relative contributions $(\sigma_{\mathrm{A}}/\sigma_{\mathrm{N}})$ 
from coherent and incoherent scattering
and the diffractive slope parameters $b_{\mathrm{N}}$ and $b_{\mathrm{A}}$
(\ref{eq:dif}) were all varied within uncertainties.
\item Background correction:  The systematic uncertainty attributable to
the background correction
was estimated by varying the $\Delta E$ cut in the analysis procedure.
\item Relative contributions of $r^{1,5,8}_{00}$ and $r^{1,5,8}_{11}$ 
to the trace Tr($r^{1,5,8}$)
((\ref{eq:Phi00}) and (\ref{eq:Phi11})) measured by the
${\cal W}(\Phi)$ distribution (\ref{eq:Phi}).
\item Possible violations of SCHC: Previous measurements~\cite{joos,chio}
covering a wide energy range have yielded non-zero values for $r^{04}_{10}$,
which can be attributed
to interference between a small spin-flip amplitude and a non-spin-flip
amplitude $|T_{\mathrm{flip}}|/|T_{\mathrm{no-flip}}|\approx0.14$
(see~Appendix~1 of Ref.~\cite{joos}).
\item Radiative effects: The spectrometer material in the path of the 
scattered positrons leaving the target is about 5\% of a 
radiation length. 
This happens to be an amount that will produce an effect similar to that 
of internal radiation,
based on the equivalent radiator approximation~\cite{motsai}.
Hence the systematic uncertainty attributable to uncorrected internal radiative
effects was estimated
by comparing fit results using Monte Carlo data generated with and without
external radiative effects. 
\end{itemize}
The uncertainty contributions thus estimated were added in quadrature.

\section{Results and Discussion}\label{sec:results}
\begin{figure}
\begin{center}
\epsfig{file=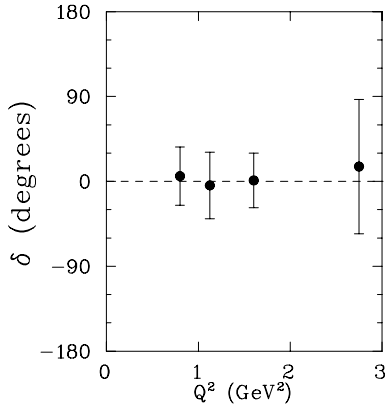,width=5cm}
\end{center}
\caption{Best-fit values for the phase difference $\delta$ between the
longitudinal and transverse amplitudes (statistical uncertainties only).}
\label{fig:delta}
\end{figure}

The best-fit values of 
the phase angle $\delta$ shown in Fig.~\ref{fig:delta}
are consistent with zero, as previously observed for
$W>3\,$GeV~\cite{bauer}, and as expected when both longitudinal
and transverse production are diffractive.

\begin{figure*}[p]\sidecaption
\epsfig{file=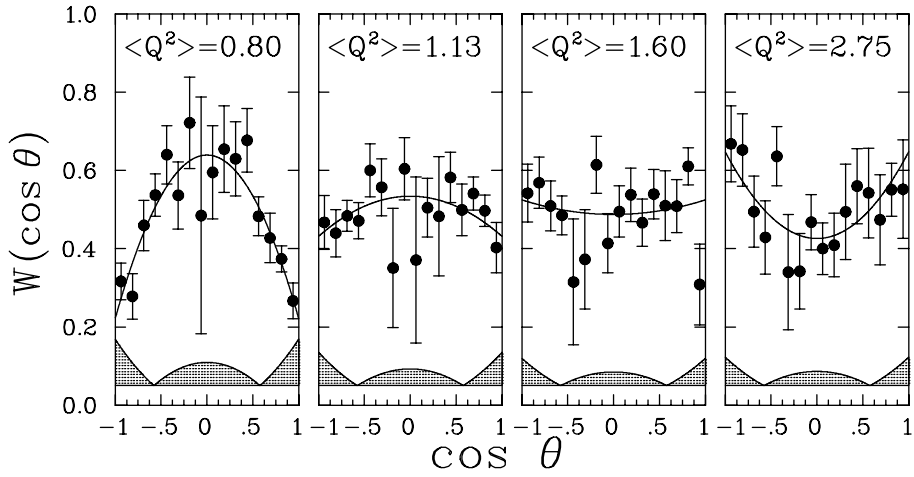,width=12cm}
\caption[dummy]{The acceptance-corrected angular distributions 
${\cal W}(\cos\theta)$ in four regions of
$Q^2$, with statistical error bars.
The solid curves are from (\protect\ref{eq:cos}) evaluated with the
best-fit matrix elements. The shaded regions indicate the
uncertainty of the curves (or of the acceptance correction to the points)
arising from the total uncertainty of
$r^{04}_{00}$ 
(see text).
The average values of $Q^2$ are shown in GeV$^2$.}
\label{fig:cos}
\end{figure*}

\begin{figure*}[p]\sidecaption
\epsfig{file=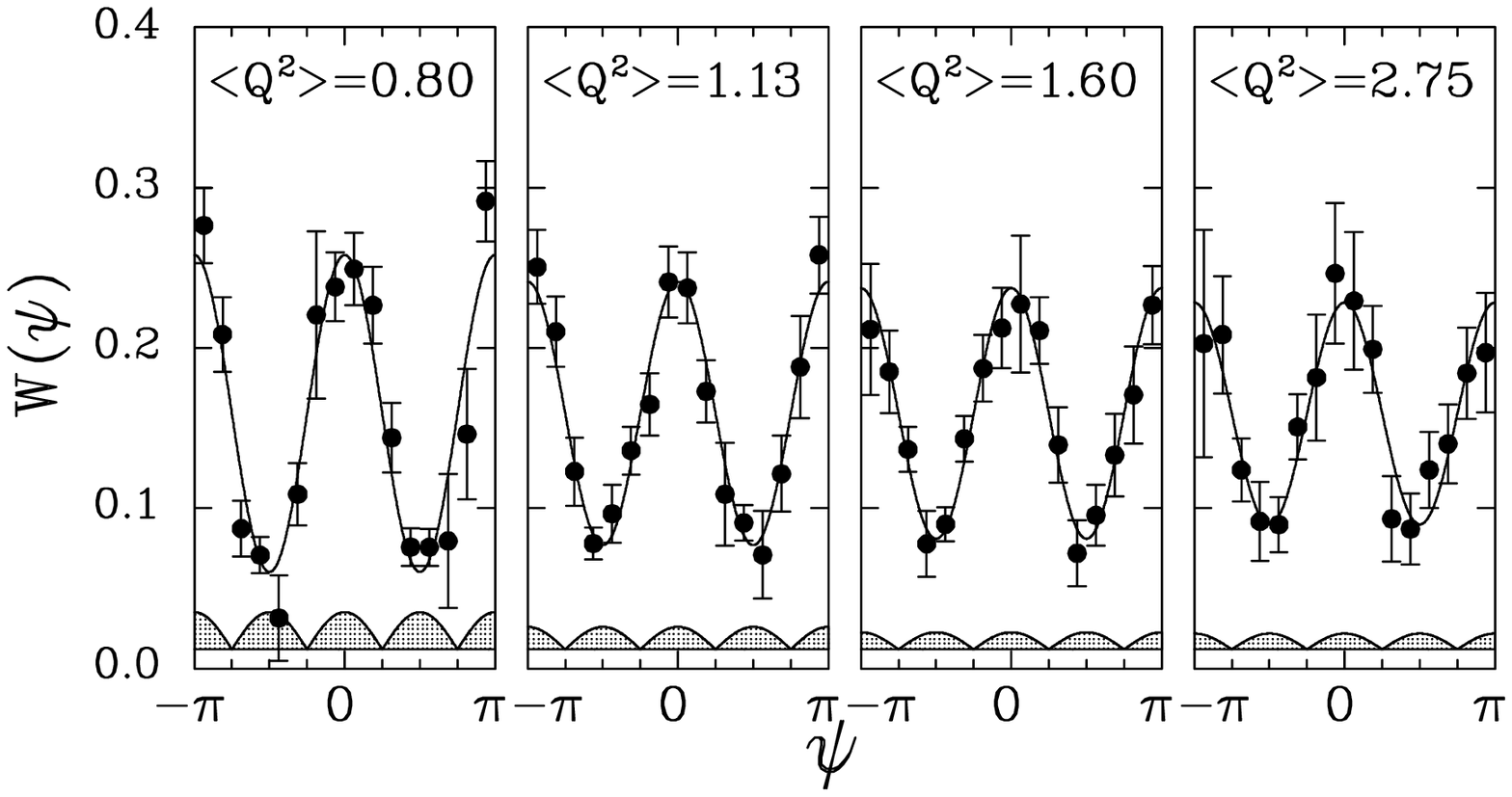,width=12cm}
\caption[dummy]{The acceptance-corrected angular distributions 
${\cal W}(\psi)$ in four regions of
$Q^2$, with statistical error bars.
The solid curves are from (\protect\ref{eq:psi}) evaluated with the
best-fit matrix elements. The shaded regions indicate the
uncertainty of the curves arising from the total uncertainty of
$r^{1}_{1-1}$.
The average values of $Q^2$ are shown in GeV$^2$.
\label{fig:psi}
}
\end{figure*}

\begin{figure*}[t]\sidecaption
\epsfig{file=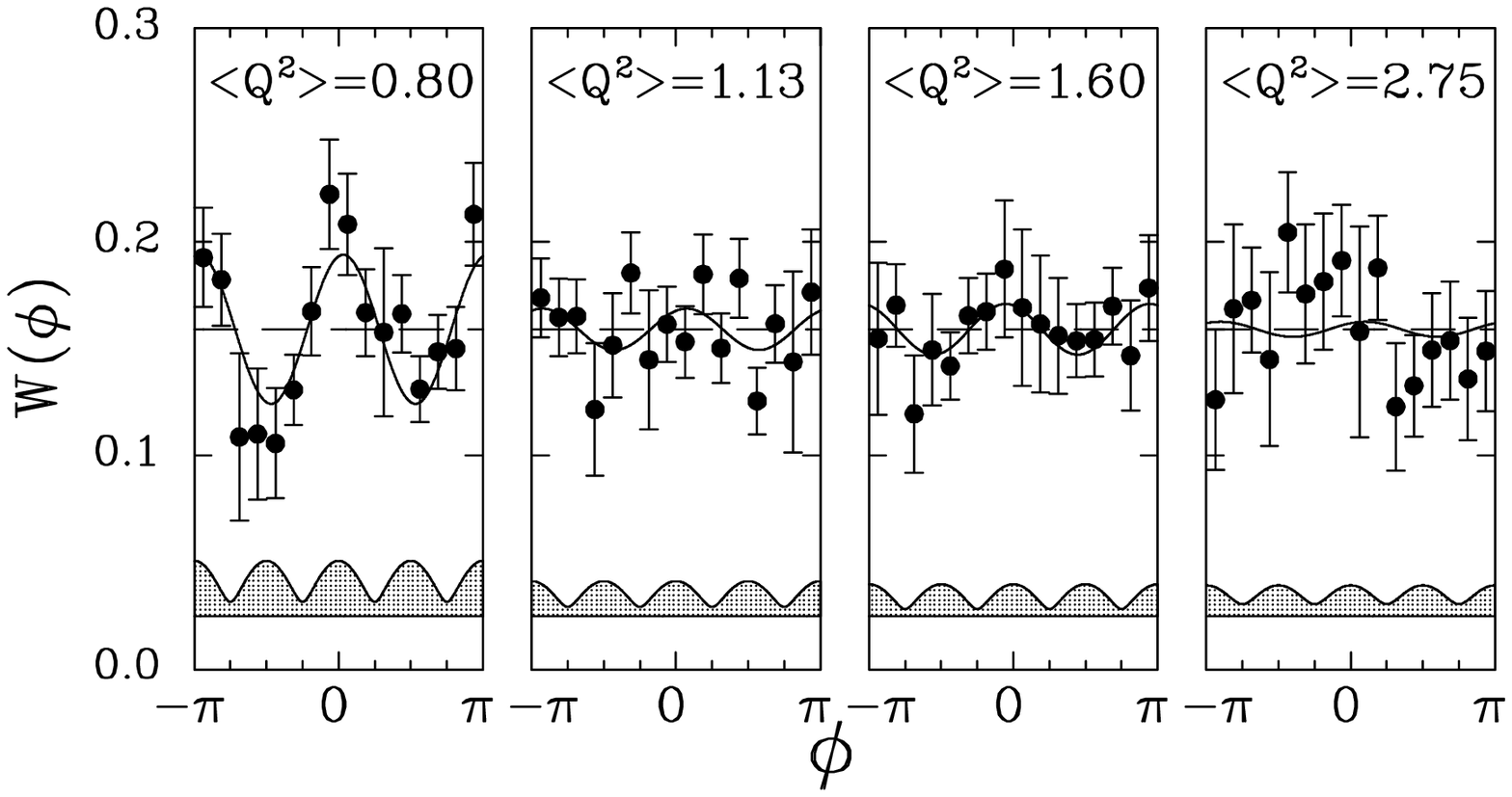,width=12cm}
\caption[dummy]{The acceptance-corrected angular distributions ${\cal W}(\phi)$ in
four regions of
$Q^2$, with statistical error bars.
The solid curves are from (\protect\ref{eq:phi}) evaluated with the
best-fit matrix elements. The shaded regions indicate the
uncertainty of the curves arising from the total uncertainties
in $r^{04}_{1-1}$ and $\Im(r^3_{1-1})$.
The average values of $Q^2$ are shown in GeV$^2$.
The dashed horizontal line represents the value $\frac{1}{2 \pi}$
predicted by SCHC.}
\label{fig:phi}
\end{figure*}

\begin{figure*}[t]\sidecaption
\epsfig{file=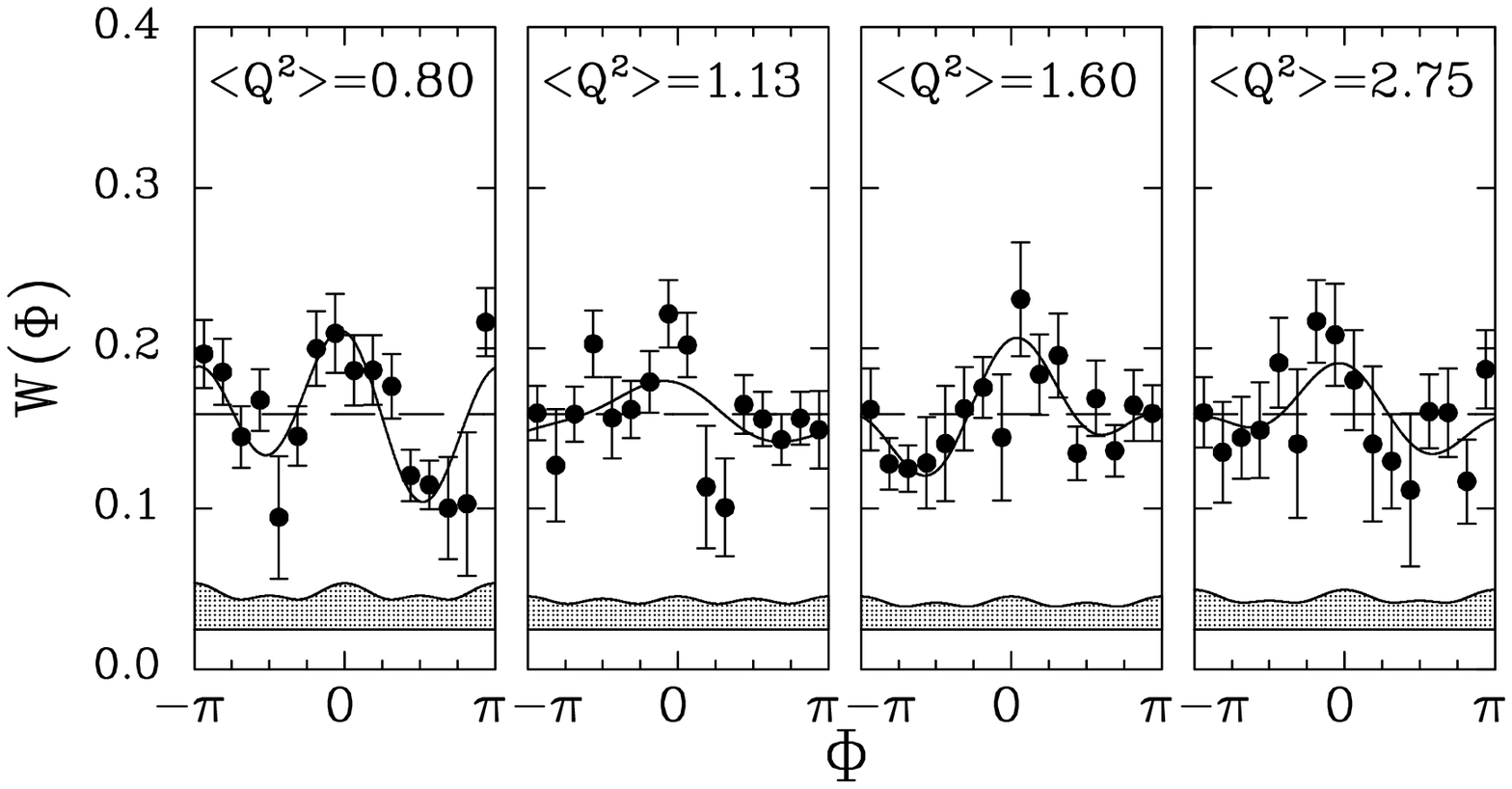,width=12cm}
\caption[dummy]{
The acceptance-corrected angular distributions ${\cal W}(\Phi)$ in four
regions of $Q^2$, with statistical error bars.
The solid curves are from (\protect\ref{eq:Phi}) evaluated with the
best-fit matrix elements.
The shaded regions indicate the
uncertainty of the curves arising from the total uncertainties
in Tr$(r^1)$, Tr$(r^5)$, and Tr$(r^8)$.
The average values of $Q^2$ are shown in GeV$^2$.
The dashed horizontal line represents the value $\frac{1}{2 \pi}$
predicted by SCHC.}
\label{fig:Phi}
\end{figure*}

Previous data have typically been presented as one-dimensional 
distributions with corrections for instrumental effects applied. 
In keeping with this tradition, an acceptance correction was trivially
obtained from the optimized matrix elements by comparing the Monte Carlo yields
($dN/d\cos\theta$,
$dN/d\phi$,
$dN/d\Phi$,
and $dN/d\psi$)
with the analytic expressions (\ref{eq:cos}--\ref{eq:psi}) for the 
angular distributions
(${\cal W}(\cos\theta)$, ${\cal W}(\phi)$, ${\cal W}(\Phi)$, and
${\cal W}(\psi)$, respectively).
The corresponding measured angular distributions after applying this correction
are shown for four values of $Q^2$
in Figs.~\ref{fig:cos},\,\ref{fig:phi},\,\ref{fig:Phi} and \ref{fig:psi}.
These figures should be interpreted with caution, as each one 
alone does not adequately represent the global fit involving 
strong correlations between the resulting matrix elements.
Systematic and statistical uncertainties combined in the
error bands apply in the same way to both the curves representing 
the results of the global fit and --- through the acceptance correction ---
the data. 

\begin{figure*}[p]\sidecaption
\epsfig{file=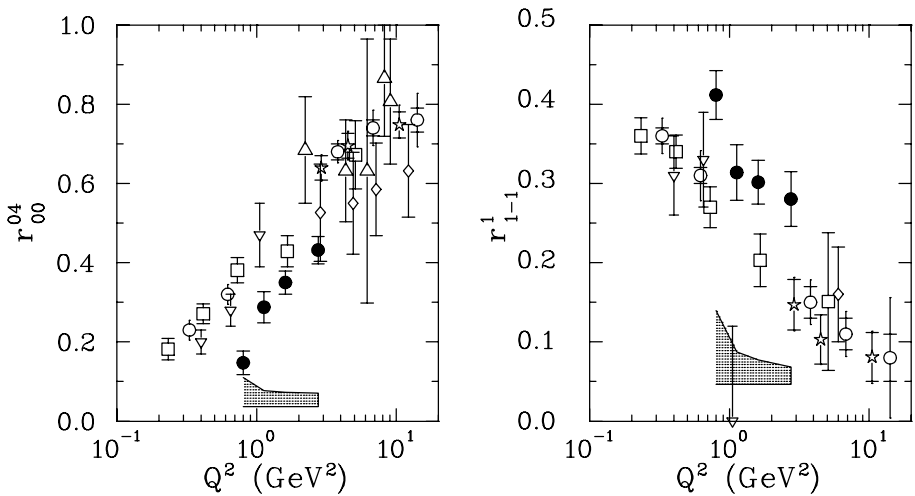,width=12cm}
\caption[dummy]{The spin density matrix elements $r^{04}_{00}$ and
$r^{1}_{1-1}$ as a function of $Q^2$ ($\bullet$), 
with the band representing the systematic uncertainty.
There are shown for comparison previous measurements 
with statistical uncertainties as (inner) error bars, from
DESY~\protect\cite{joos}~($\triangledown$),
EMC~\protect\cite{emc}~($\triangle$),
NMC~\protect\cite{nmc}~($\diamond$),
E665~\protect\cite{e665}~($\Box$),
ZEUS~\protect\cite{zeus}~($\circ$),
and
H1~\protect\cite{h1}~($\star$).
Where the systematic uncertainties were specified~\protect\cite{zeus,h1}, 
they are added in quadrature to produce the outer error bars. 
}
\label{fig:plot}
\end{figure*}
\begin{table*}[p]\sidecaption
\begin{tabular}{ccccc}
\hline\noalign{\smallskip}
$\langle Q^2 \rangle$&$\langle \epsilon \rangle$ & measured $r^{04}_{00}$
& measured $r^1_{1-1}$ & $ r^{1}_{1-1}$(SCHC)\\
 (GeV$^2)$ & & & & from $r^{04}_{00}$ \\
\noalign{\smallskip}\hline\noalign{\smallskip}
  0.80&0.76& $0.147\pm0.030\pm0.073 $ & $ 0.412\pm0.031\pm0.093 $ & $ 0.43\pm 0.02 $\\
  1.13&0.82& $0.288\pm0.040\pm0.041 $ & $ 0.314\pm0.035\pm0.041 $ & $ 0.36\pm 0.02 $\\
  1.60&0.81& $0.350\pm0.029\pm0.036 $ & $ 0.302\pm0.028\pm0.031 $ & $ 0.32\pm 0.02 $\\
  2.75&0.78& $0.432\pm0.035\pm0.034 $ & $ 0.280\pm0.034\pm0.022 $ & $ 0.28\pm 0.02 $\\
\noalign{\smallskip}\hline
\end{tabular}
\caption[dummy]{Spin density matrix elements
 describing the distributions in decay angle ${\cal W}(\cos\theta)$
 and in polarization angle ${\cal W}(\psi)$.
 The measured values of $r^{1}_{1-1}$ are compared with the predictions
 based on (\protect\ref{eq:np}) using the measured values of $r^{04}_{00}$.
 The first (second) uncertainty is statistical (systematic).
\vskip -1.25 cm           
}
\label{tab:2}
 \end{table*}

\begin{figure*}[t]\sidecaption
\epsfig{file=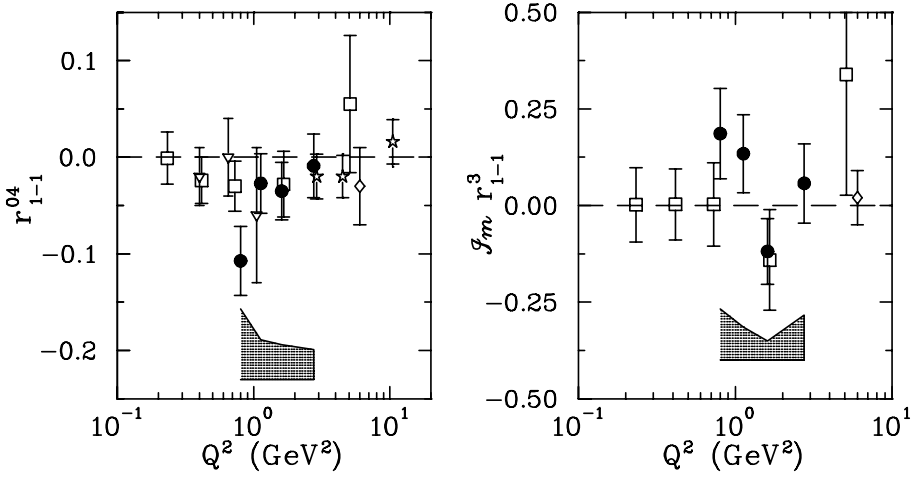,width=12cm}
\caption[dummy]{The spin density matrix elements for $r^{04}_{1-1}$ and
$\Im(r^{3}_{1-1})$ as a function of $Q^2$ ($\bullet$),
with the band representing the systematic uncertainty.
There are shown for comparison previous measurements 
with statistical uncertainties only
from
DESY~\protect\cite{joos}~($\triangledown$), 
NMC~\protect\cite{nmc}~($\diamond$), 
E665~\protect\cite{e665}~($\Box$), and from
H1~\protect\cite{h1}~($\star$) where the systematic uncertainties are
specified to be negligible in this case.
}
\label{fig:plot2}
\end{figure*}
 \begin{table*}[t]\sidecaption
 \begin{tabular}{cccc}
\hline\noalign{\smallskip}
$\langle Q^2 \rangle$ & $\langle \epsilon \rangle$ &
$r^{04}_{1-1}$ & $\Im(r^3_{1-1})$ \\
 (GeV$^2)$ & & & \\
\noalign{\smallskip}\hline\noalign{\smallskip}
  0.80&0.76& $-0.107\pm0.036\pm0.073 $& $ 0.186\pm0.117\pm0.132 $\\
  1.13&0.82& $-0.027\pm0.031\pm0.041 $& $ 0.134\pm0.101\pm0.086 $\\
  1.60&0.81& $-0.035\pm0.030\pm0.036 $& $-0.119\pm0.085\pm0.049 $\\
  2.75&0.78& $-0.009\pm0.033\pm0.034 $& $ 0.057\pm0.103\pm0.116 $\\
\noalign{\smallskip}\hline
\end{tabular}
 \caption{Spin density matrix elements
 describing the the azimuthal decay-angle distribution ${\cal W}(\phi)$.
 The first (second) uncertainty is statistical (systematic).}
 \label{tab:3}
 \end{table*}

\subsection{Matrix Elements Allowed by SCHC}

As observed in previous measurements, the ${\cal W}(\cos\theta)$
distribution in Fig.~\ref{fig:cos} shows a dramatic shift as $Q^2$ increases,
from a predominantly $\sin^2{\theta}$ shape to
a predominantly $\cos^2{\theta}$ shape, indicating a transition
from transverse to longitudinal production (see (\ref{eq:cos})).
The structure of ${\cal W}(\psi)$ visible in Fig.~\ref{fig:psi}
is entirely due to transverse $\rho^0$ mesons ($r^1_{1-1}$ in (\ref{eq:psi})),
so that as the longitudinal component increases, the `damping' of 
the ${\cal W}(\psi)$ distribution indicates a complementary decrease 
in $r^1_{1-1}$ because of the dilution from longitudinal $\rho^0$ mesons.
The measured matrix elements
$r^{04}_{00}$ and $r^1_{1-1}$ 
are plotted versus $Q^2$ in Fig.~\ref{fig:plot} 
and are listed in Table~\ref{tab:2}.
Comparison with data of other experiments in Fig.~\ref{fig:plot} is 
complicated by their different values of the polarization
parameter $\epsilon$.
For that reason, this issue will be revisited when discussing the
values of $R$ derived from the data in Fig.~\ref{fig:plot}.

\begin{figure*}[t]\sidecaption
\epsfig{file=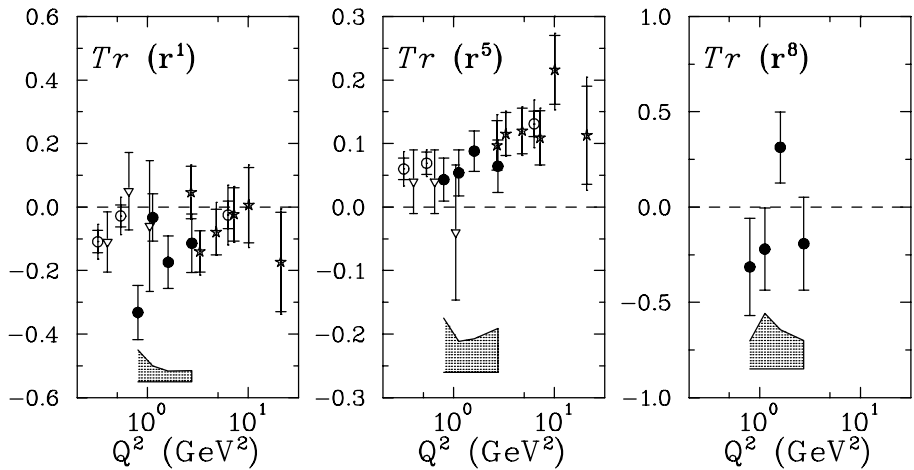,width=12cm}
\caption[dummy]{The matrix elements
Tr$(r^1)$,
Tr$(r^5)$ and
Tr$(r^8)$,
as a function of $Q^2$ ($\bullet$),
with the band representing the systematic uncertainty.
There are also shown previous measurements from 
H1~\protect\cite{h1}~($\star$), and also values from
DESY~\protect\cite{joos}~($\triangledown$) 
(with statistical uncertainties only)
and ZEUS~\protect\cite{zeusvio}~($\circ$), calculated 
from measurements of $r^{1,5}_{11}$ and $r^{1,5}_{00}$.
Where the systematic uncertainties were specified~\protect\cite{zeus,h1}, 
they are added in quadrature to produce the outer error bars. 
}
\label{fig:plot3}
\end{figure*}
\begin{table*}[t]\sidecaption
\begin{tabular}{ccccc}
\hline\noalign{\smallskip}
$\langle Q^2 \rangle$&$\langle \epsilon \rangle$ &
Tr($r^1$) & Tr($r^5$) &Tr($r^8$)\\
 (GeV$^2)$ & & & & \\
\noalign{\smallskip}\hline\noalign{\smallskip}
  0.80&0.76& $-0.332\pm0.085\pm0.100 $& $ 0.043\pm0.034\pm0.085 $& $-0.314\pm0.256\pm0.148 $\\
  1.13&0.82& $-0.033\pm0.075\pm0.049 $& $ 0.054\pm0.036\pm0.048 $& $-0.221\pm0.215\pm0.291 $\\
  1.60&0.81& $-0.174\pm0.083\pm0.033 $& $ 0.088\pm0.032\pm0.052 $& $ 0.314\pm0.186\pm0.206 $\\
  2.75&0.78& $-0.114\pm0.092\pm0.035 $& $ 0.064\pm0.041\pm0.069 $& $-0.192\pm0.243\pm0.150 $\\
\noalign{\smallskip}\hline
\end{tabular}
 \caption{Spin density matrix elements
 describing the production angle distributions ${\cal W}(\Phi)$.
The first (second)
uncertainty is statistical (systematic).
\vskip -1.1 cm      
\label{tab:4}
}
 \end{table*}

If SCHC is valid and the $\gamma^* N\rightarrow\rho^0 N$ reaction 
proceeds via the $t$-channel exchange of a particle with natural parity
(${\mathrm{J}}^P=0^+$, $1^-$,...), then the tensor alignment of the
mesons in the
scattering plane is related to the transverse fraction of mesons by
\begin{equation}
2r^{1}_{1-1}=(1-r^{04}_{00}).
\label{eq:np}
\end{equation}
Thus the decay pions from transverse $\rho^0$ mesons
are preferentially emitted in the scattering plane, while those
emitted from longitudinal $\rho^0$ mesons show no $\psi$ dependence.
A comparison of the last two columns of Table~\ref{tab:2} reveals that
the measured values of $r^{04}_{00}$ and $r^{1}_{1-1}$ 
are in excellent agreement with (\ref{eq:np}), with a 
total $\chi^2$ of 2.7 for 4 degrees of freedom, accounting for correlations.
This demonstrates that the
reaction is dominated by helicity-conserving amplitudes with natural
parity exchange.

\subsection{Matrix Elements forbidden by SCHC}

The distributions in the azimuthal decay and production 
angles $\phi$ and $\Phi$ are shown in Figs.~\ref{fig:phi} and ~\ref{fig:Phi};
the matrix elements describing the distributions
can be found in Tables~\ref{tab:3} and~\ref{tab:4} respectively,
and are plotted versus $Q^2$ in Figs.~\ref{fig:plot2} and \ref{fig:plot3}.
Where other data exist, they are consistent. The present data 
for Tr$(r^1)$ and Tr$(r^5)$ 
interpolate in $Q^2$ between the photo- and lepto-production data
from the collider experiments~\cite{zeusvio,h1}, but appear consistent
in spite of the large difference in $W$.
No previous data for Tr$(r^8)$ exist.
The extraction of $\Im(r^3_{1-1})$ and Tr$(r^8)$ depends on the use of
a polarized lepton beam.
Even though a single beam helicity was used, the effect of the
term involving $r^3_{1-1}$ is not visible, and
only a $\cos 2\phi$ variation can be seen in ${\cal W}(\phi)$.
This dependence is clearly visible in Fig.~\ref{fig:phi} at the
smallest value of $Q^2$, although this variation has only marginal significance
in comparison with the total uncertainties.
Within the systematic uncertainty of the measurement, the present
results are in fair agreement with the SCHC prediction that
$r^{04}_{1-1}=$ $\Im(r^3_{1-1})=$ Tr$(r^1)=$ Tr$(r^5)=$ Tr$(r^8)\equiv 0$,
though small negative values for $r^{04}_{1-1}$ and Tr($r^1$) 
(for the latter 1.8~standard deviations at the smallest $Q^2$, 
and about 2.5~$\sigma$ averaged over $Q^2$) 
could both be attributed to the interference of a small double-spin-flip
amplitude with a non-spin-flip amplitude.
Similarly, the non-zero value of
Tr($r^5$) --- 3.5~standard deviations averaged over $Q^2$, ignoring the 
systematic uncertainties which only in this case are dominated by 
radiative effects --- could be caused by interference of a spin-flip amplitude
with a helicity conserving amplitude
(see~Appendix A of~\cite{wolf} and Appendix 1 of~\cite{joos}).
The present data are in agreement with the 
significantly positive values for Tr$(r^5)$ from the 
HERA colliding-beam experiments at large $W$, which reflect the similar 
SCHC-violating behaviour reported for their 
values of $r^5_{00}$~\cite{zeusvio,h1}. However, 
radiative effects are not included in their systematic uncertainties.
This violation by $r^5_{00}$ was not observed in measurements with
apparently similar precision at low $W (\sim 2.3\,$GeV)~\cite{joos}. 

\subsection{Ratio of Longitudinal to Transverse Cross Sections}
With the assumption of $s$-channel helicity
conservation,
knowledge of the virtual photon
polarization allows extraction of the ratio of the longitudinal
to transverse cross sections for diffractive $\rho^0$ production 
\begin{equation}
R\equiv {{\sigma_L}\over{\sigma_T}}=
{{1}\over{\epsilon}} {{r^{04}_{00}}\over{1-r^{04}_{00}}}
\label{eq:rlt}
\end{equation}
without having to combine data at different beam energies.
The right side of (\ref{eq:rlt}) can
be interpreted simply as the fraction $r^{04}_{00}$ of longitudinal
$\rho^0$ mesons normalized to the fraction $\epsilon$ of
longitudinal photons, 
divided by the fraction $(1-r^{04}_{00})$ of transverse
$\rho^0$ mesons normalized to the unit flux of transverse photons.

As $r^{04}_{00}$ in (\ref{eq:rlt}) depends
only on the squares of amplitudes,
(\ref{eq:rlt}) is robust against SCHC violations. 
For instance, the SCHC-violating spin-flip amplitudes reported
in~\cite{joos,chio} to be as large as 14\% of the SCHC-conserving non-spin-flip
amplitude require only a 2\% correction after squaring them.
Moreover, the double-spin-flip amplitudes
$T_{-11}$ and $T_{1-1}$ couple transverse
$m=\pm 1$ photons to $m=\mp 1$ $\rho^0$ mesons,
and therefore do not affect the validity of (\ref{eq:rlt}).

\begin{figure}
\begin{center}
\epsfig{file=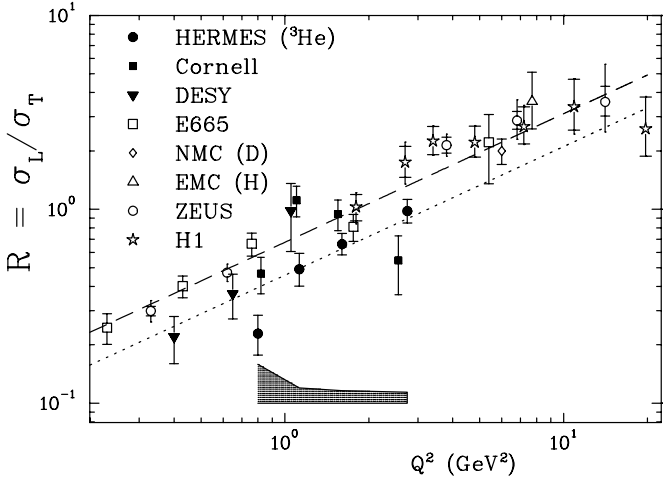,width=8.7cm}
\end{center}
\caption[dummy]{The ratio $R\equiv\sigma_L/\sigma_T$ determined from the measured
values of $r^{04}_{00}$ using (\protect\ref{eq:rlt}) ($\bullet$).
The errors bars are statistical only, while
the shaded region indicates the systematic uncertainty of the present data.
Previous measurements are shown for comparison, with experiment names
as in Fig.~\protect\ref{fig:plot} with the addition of Cornell for
ref.~\protect\cite{cassel}. 
Where the systematic uncertainties were specified~\protect\cite{zeus,h1}, 
they are added in quadrature to produce the outer error bars. 
The dashed (dotted) line
represents a simultaneous fit according to
(\protect\ref{eq:powerlaw}) to data with average $W>4$ and 
above (below) $7\,$ GeV,
which are shown as open (filled) data points.
The DESY and Cornell data for $2<W<4$\,GeV are also plotted but were not
included in the fit (see text).
} 
\label{fig:rlt}
\end{figure}

Equation~(\ref{eq:rlt}) was used to determine $R$; the results are
listed in Table~\ref{tab:5}, and plotted in Fig.~\ref{fig:rlt} 
together with previous data with relevant precision in the regime $W>4\,$GeV.
Gauge invariance restricts $R$ to be zero at $Q^2=0$, and dimensional
arguments imply that the longitudinal cross section dominates at 
asymptotically large $Q^2$~\cite{pqcd3}.
All of the sets of existing data are consistent with a monotonic interpolation
between these extremes, and show
that both the transverse and longitudinal production mechanisms are important
at the energy and $Q^2$ accessible at \HERMES. 

The constraint $R(Q^2=0)=0$ and the prediction by several models
that the longitudinal and transverse cross sections should differ by
some power of $Q^2$~\cite{bauer,nonper1,pqcd3} suggest a fit of the
form
\begin{equation}\label{eq:powerlaw}
R(Q^2)=c_0(W)\ \left[\frac{Q^2}{M^2_{\rho}}\right]^{c_1}.
\end{equation}
The available data now span a wide range of centre-of-mass energies
both above and below $W =7\,$GeV. 
This is close to the energy at which the $W$-dependence of the
$\rho^0$ production cross section makes a transition from a decreasing behavior
attributed to Reggeon exchange to a $W^{0.22}$ dependence attributed
to Pomeron exchange.
The improved precision that the present data add to the lower
energy region allows
the fit to be done with two values of $c_0$~\cite{sakurai}:
one for the data from high energy muon beams and collider experiments
spanning  $7 \lesssim W<140\,$GeV (EMC~\cite{emc}, NMC~\cite{nmc}, 
E665~\cite{e665}, ZEUS~\cite{zeus}, and H1~\cite{h1}),
and another for the present data in the lower beam energy range
$3.8<W<6.5\,$GeV,
yielding
\begin{eqnarray}
\nonumber
c_0
& = & \left\{ \begin{array}{ll}
        0.32\pm 0.04,   & \hspace{4mm} 4< \langle W\rangle < 7\rm\,\,GeV \\
        0.48\pm 0.03,   & \hspace{10.5mm} \langle W\rangle > 7\rm\,\,GeV \\
                 \end{array}
          \right. \\
\nonumber
c_1 & = & 0.66\pm 0.03.
\end{eqnarray}
The $\chi^2$ per degree of freedom for the combined fit is 20.7/20,
calculated from the statistical uncertainties, 
plus the systematic uncertainties in quadrature where the latter are available.
Difficulties in assessing the systematic uncertainties
in some of the data sets preclude a comprehensive treatment of
all the systematic uncertainties.

\begin{table}
\begin{center}
\caption{$R$ as determined from $r^{04}_{00}$.}
\label{tab:5}
\begin{tabular}{ccc}
\hline\noalign{\smallskip}
$\langle Q^2 \rangle$&$\langle \epsilon \rangle$ & $R$ \\
(GeV$^2)$ & & \\
\noalign{\smallskip}\hline\noalign{\smallskip}
0.80&0.76& $0.23^{+0.06+0.15}_{-0.05-0.12} $\\ 
\noalign{\smallskip}
1.13&0.82& $0.49^{+0.10+0.10}_{-0.09-0.09} $\\ 
\noalign{\smallskip}
1.60&0.81& $0.66^{+0.09+0.11}_{-0.08-0.10} $\\ 
\noalign{\smallskip}
2.75&0.78& $0.98^{+0.15+0.14}_{-0.13-0.13} $\\
\noalign{\smallskip}\hline
\end{tabular}
\end{center}
\end{table}

The only previous $R$ data~\cite{dakin} in the \HERMES\ energy region 
is consistent but insufficiently precise to influence the fit. 
Inclusion of either or both
of two existing data sets~\cite{joos,cassel} in the range $2<W<4\,$GeV 
changes the results of the fit by less than one standard deviation,
and does not significantly improve the precision. (It may be noted
that the data of reference~\cite{cassel} are not well-described by the fit
function --- their $\chi^2$ contribution would be
similar to that of all of the rest of the fitted data in both 
energy regimes.)

As observed in previous experiments, the $Q^2$ dependence of $R$
is in marked disagreement with the prediction of either VMD~\cite{sakurai}
or pQCD~\cite{pqcd3} that $c_1\equiv 1$.
The value of $c_0$ is found to increase with centre-of-mass energy
in the range above $W=4\,$GeV where both longitudinal and transverse
production are clearly diffractive.\footnote{
Experiments at lower energy~\cite{joos,chio} provide evidence that 
for fixed $Q^2$, $R$ increases also as $W$ falls below 2\,GeV.} 
It has been shown~\cite{frankfurt} that the energy dependences
of the longitudinal and transverse couplings for gluon exchange
are identical, so that $R$ is independent of energy at high energy;
however,
recent theoretical work~\cite{guidal} has suggested that at lower
energy where quark exchange is dominant,
there may be an additional class of diagrams that enhance the
transverse couplings. The observed energy dependence of $c_0$ 
is consistent with this trend.
However, it should be kept in mind that the lower energy fit 
to the present data set on the $^3$He target could
be influenced by nuclear effects, although it has been suggested
that this would {\em increase} $R$~\cite{bauer}.
Other matrix elements accessible with a real photon beam were found
to be the same for coherent $\rho^0$ production on deuterium
and incoherent production on either nucleon in deuterium~\cite{eisen}.
Furthermore, those data are consistent with other similar measurements
on hydrogen~\cite{ballam}. Finally, coherent production on such a light
nucleus as $^3$He accounts for only about 25\% of the events (see 
Fig.~\ref{fig:tprime}).
Hence a nuclear effect is unlikely to account for the observed
difference from the results at higher energy.\footnote{
The distinction between coherent and incoherent processes can be
better studied in a heavier nucleus such as $^{14}$N. An analysis
of such data is underway.}

\section{Summary}\label{sec:sum}
  The production and decay-angle distributions for the combined coherent 
and incoherent polarized electroproduction
of $\rho^0$ mesons 
from $^3$He
have been measured
at photon virtuality $0.5<Q^2<4\,$GeV$^2$ and photon-nucleon 
centre-of-mass energy
$3.8<W<6.5\,$GeV.
The spin-density matrix elements
$r^{04}_{00}$,
$r^{04}_{1-1}$, 
$r^1_{1-1}$, 
$r^{3}_{1-1}$, 
${\mathrm{Tr}}\,(r^1)$,
${\mathrm{Tr}}\,(r^5)$ and
${\mathrm{Tr}}\,(r^8)$
are found to be mostly consistent with s-channel helicity conservation
and natural parity exchange in the $t$-channel.
Possible deviations from SCHC at the few sigma level were found for Tr($r^1$).
The ratio $R$ of the cross sections for
longitudinal and transverse photons was determined from
the decay-angle distribution over a range $0.5<Q^2<4\,$GeV$^2$
and at $\langle W\rangle = 5\,$GeV.
The $Q^2$ dependence
of $R$ is consistent with previous experiments, exhibiting
a monotonic increase with $Q^2$; however, the present data 
now suggest that 
at a given $Q^2$ value, $R$ increases with 
average photon-nucleon centre-of-mass energy for such energies above 4\,GeV,
assuming that nuclear effects in coherent
production on $^3$He can be neglected.

\begin{acknowledgement}
We gratefully acknowledge the DESY Directorate for its support and 
the DESY staff and the staffs of the collaborating institutions. 
We particularly appreciate the efforts of the HERA machine group
in providing high beam polarization.
Additional support for this work was provided by
the Deutscher Akademischer Austauschdienst (DAAD) and 
INTAS, HCM,  and TMR network contributions from the European Community.
\end{acknowledgement}

\end{document}